%% file: links-gtopdb.tex
\documentclass[a4paper, 11pt]{article}
\input{preamble}

\title{Cross-tier web programming for curated databases: \\A case study}
\author{Simon Fowler\\University of Glasgow \and
Simon D.\ Harding\\University of Edinburgh \and
Joanna Sharman\\Novo Nordisk \and
James Cheney\\University of Edinburgh}

\begin{document}
\maketitle
\begin{abstract}
Curated databases have become important sources of information across several scientific
disciplines, and as the result of manual work of experts, often become important reference
works. Features such as provenance tracking, archiving, and data citation are widely regarded as
important features for the curated databases, but implementing such features is
challenging, and small database projects often lack the resources to do so.

A scientific database application is not just the relational database itself,
but also an ecosystem of web applications to display the data, and applications
which allow data curation. Supporting advanced curation features requires
changing all of these components, and there is currently no way to provide such
capabilities in a reusable way.

Cross-tier programming languages allow developers to write a web application in
a single, uniform language. Consequently, database queries and updates can be
written in the same language as the rest of the program, and it should be
possible to provide curation features via program transformations. As a
step towards this goal, it is important to establish that realistic
curated databases can be implemented in a cross-tier programming language.

In this article, we describe such a case study: reimplementing the
web frontend of a real- world scientific database, the IUPHAR/BPS
Guide to PHARMACOLOGY (GtoPdb), in the Links cross-tier programming
language.
We show how programming language features such as
language-integrated query simplify the development process, and rule
out common errors.
Through an automated functional correctness evaluation, we show that the Links
implementation correctly implements the functionality of the official version.
Through a comparative performance evaluation, we
show that the Links implementation performs fewer database queries,
while the time needed to handle the queries is comparable to the official
Java version.  Furthermore, while there is some overhead to using Links because
of its comparative immaturity compared to Java, the Links version is usable as a
proof-of-concept case study of cross-tier programming for curated databases.
\end{abstract}

\section{Introduction}
Curated databases have become important data resources across several scientific
disciplines.  Such databases collect the current state of knowledge about a
topic and many have become important reference works.  They are constructed
through the manual effort of experts, often over a long timespan, and it is
widely appreciated that versioning and provenance-tracking are important for
assessing the validity and freshness of the data, or tracing the origin of
errors or discrepancies~\citep{BunemanCTV08:curated}.  Unfortunately, implementing
support for fine-grained provenance-tracking or versioning is a challenging
task, usually performed on a system-by-system basis.  Many such
database projects, particularly smaller or shorter-term ones, lack the
resources and expertise to do this.

A curated database is not just an isolated relational
database, but also has surrounding infrastructure such as a web application
for viewing or searching the data, and an editing interface used by the database
curators to add or modify data.  Each of these components are nontrivial to
develop.  A typical web application is really a distributed program involving
code running on several “tiers”: Java or Python running on the server,
JavaScript and HTML on the web browser, and SQL on a database.  Curation
interfaces can be either web applications or traditional client-server database
applications; in either case, modifying such a system is a nontrivial task,
especially when the added functionality spans two or all three tiers.

Thus, augmenting an existing curated database web application (or designing a new system
from scratch) to provide features such as versioning,
provenance-tracking, or citation requires taking these requirements
into account across two or more system layers, adding complexity
beyond that of the basic functionality of the system.  General-purpose techniques have been explored for supporting
such features~\citep{BunemanCTV08:curated,BunemanKTT04:archiving}, but
there is currently no way to provide them in a reusable way.

\paragraph{Cross-tier web programming.}
Cross-tier programming languages~\citep{CooperLWY06:links,RadannePVB16:eliom,SerranoGL06:hop,Chlipala15:urweb} have been proposed to simplify web and database
application programming.  The vision is that the programmer should only need to
write a single program in a single language; the language implementation then
takes care of the details of partitioning the program into client, server and
database components, distributing the code, and coordinating their communication
in the running program.
A major benefit is the fact that database queries and
updates can be written and checked for consistency in the same language as the
rest of the program.  In principle, advanced features such as
provenance and versioning could be provided for such programs by
program transformation: that is, by rewriting the program (possibly
with some lightweight annotations) so that new functionality is
implemented according to a high-level pattern.
Indeed,~\citet{FehrenbachC18:provenance} have already shown how provenance
tracking can be added as a programming language feature: a user must simply
change a keyword in a query in order to obtain provenance metadata, rather than
hand-crafting provenance tracking per application.

We argue that cross-tier programming is well-suited for curated databases: by
using a single cross-tier language rather than a conventional multi-language
approach, curated database developers should be able to focus on their
application logic, and could (in the future) use pre-packaged techniques for
provenance-tracking and archiving provided by the language implementation (or
even a library).

However, to date, cross-tier programming languages have not been
widely used for curated databases.
Before investigating the use of cross-tier programming languages to provide
language-integrated curatorial support, we must first ask:

\begin{quote}
  Are cross-tier programming languages capable of implementing \emph{realistic}
  curated databases?
\end{quote}

\paragraph{Contributions.}
In this article, we answer the above question in the affirmative. We provide the
first case study of cross-tier web programming for scientific databases by using
Links, a functional, cross-tier web programming
language~\citep{CooperLWY06:links}, to implement a workalike web front-end for
the IUPHAR/BPS Guide to PHARMACOLOGY Database (GtoPdb), an important curated
pharmacological database~\citep{Armstrong20:gtopdb}. Links is a research project
that has been developed in Edinburgh over many years, and is not a widely-used
mainstream programming language; however, by using it to develop case studies
such as this one, we plan to demonstrate the value of cross-tier programming for
scientific databases, and evidence the viability of language-based support for
curation.

In the remainder of the article, we describe the background of GtoPdb and why it
is an interesting database to use as a case study, and describe some aspects of
the Links implementation.
We then report on the evaluation of the case study, showing broad
functional equivalence via an automated functional correctness analysis,
and report on the results of a performance evaluation. The performance
evaluation shows that the Links implementation performs comparably with the
official Java implementation, as well as providing lower overall query counts
and more predictable performance results for database queries.
We conclude with a summary of lessons learned so far and directions for future
work.

\section{Background}

The International Union of Basic and Clinical Pharmacology (IUPHAR) / British
Pharmacological Society (BPS) Guide to PHARMACOLOGY (GtoPdb) is an
expert-curated database that captures interactions between human proteins
(``targets'') and ligand molecules from the pharmacological and medicinal
chemistry literature. The resource is open-access and intended as a ``one-stop
shop'' portal to pharmacological information. It provides a searchable database
with quantitative information on 3,000 drug targets and related proteins,
organised into families, and 9,700 approved and investigational drugs,
antibodies, and natural hormones, metabolites, and neurotransmitters that act on
them. GtoPdb provides succinct overviews, key references and recommended
experimental ligands for each target. It is a useful resource for researchers
and students in pharmacology and drug discovery and provides the general public
with accurate information on the basic science underlying drug action.

GtoPdb has its origin in IUPHAR-DB which was first compiled in
2003~\citep{Harmar09:iuphar, Sharman11:iuphar, Sharman13:iuphar}. Its scope was expanded between
2012 and 2015 to define the data-supported druggable genome, and was renamed
GtoPdb~\citep{Pawson14:gtopdb, Southan16:gtopdb, Armstrong20:gtopdb}.  GtoPdb is distinguished by a
unique model of data collection and curation, with the guidance and support of
the Nomenclature Committee of IUPHAR (NC-IUPHAR) and its 96 target class
subcommittees. These subcommittees comprise over 500 pharmacology experts who
provide regular updates and contributions to GtoPdb.

The GtoPdb web application communicates with two underlying databases: a main,
PostgreSQL, database which contains the bulk of the data and a second, Oracle,
database which contains ligand structure information. The main application layer
is written in Java, with static pages written in JavaServer Pages (JSP), JavaScript and HTML. The
curation interface (i.e., the interface used by curators to create and edit the
database) is a custom-built, standalone Java application with a GUI.

GtoPdb is a substantial curated database: the 2019 release comprises 89
megabytes of data contained in 181 tables. As a measure of scope, the Java
codebase for the web interface (which includes some pages out of the scope of
our reimplementation) stands at 17935 lines of code for data transformation
code; 28819 lines of JSP rendering code; and a data access
layer (which also contains query code used for the curation interface)
consisting of 43129 lines of code, written over a period of 16 years.

\section{Reimplementing GtoPdb in Links}

\begin{figure}
  \begin{center}
  \begin{subfigure}{0.45\textwidth}
    \includegraphics[width=\textwidth]{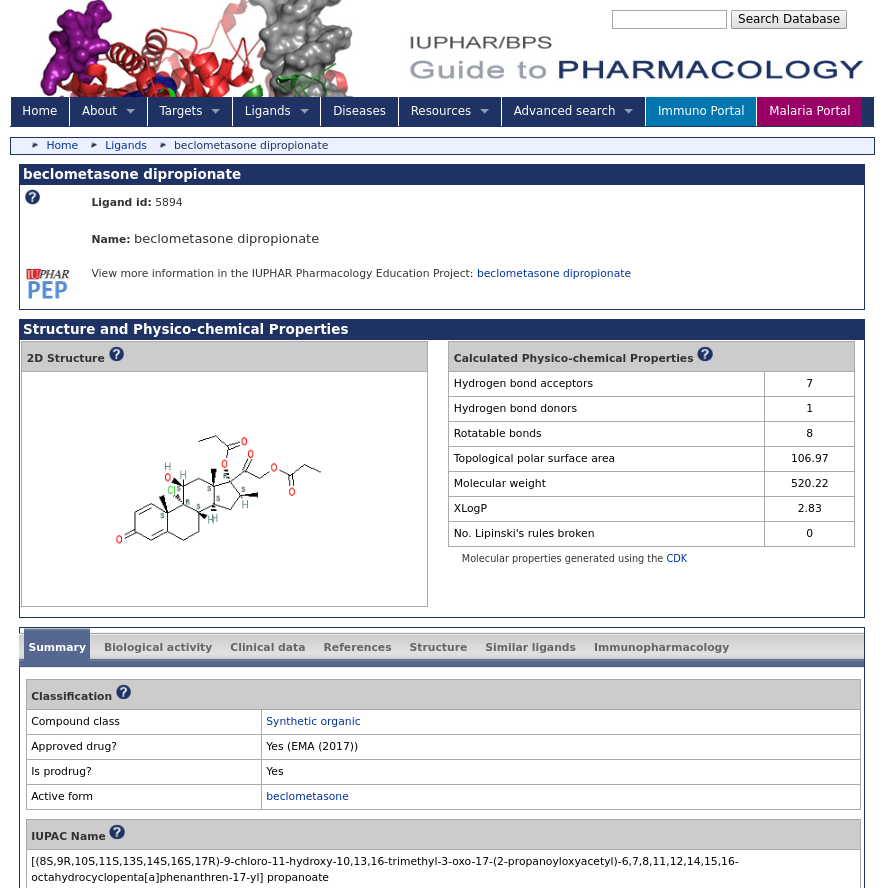}
    \caption{Official GtoPdb}
    \label{fig:screenshot:official}
  \end{subfigure}
  \qquad
  \begin{subfigure}{0.45\textwidth}
    \includegraphics[width=\textwidth]{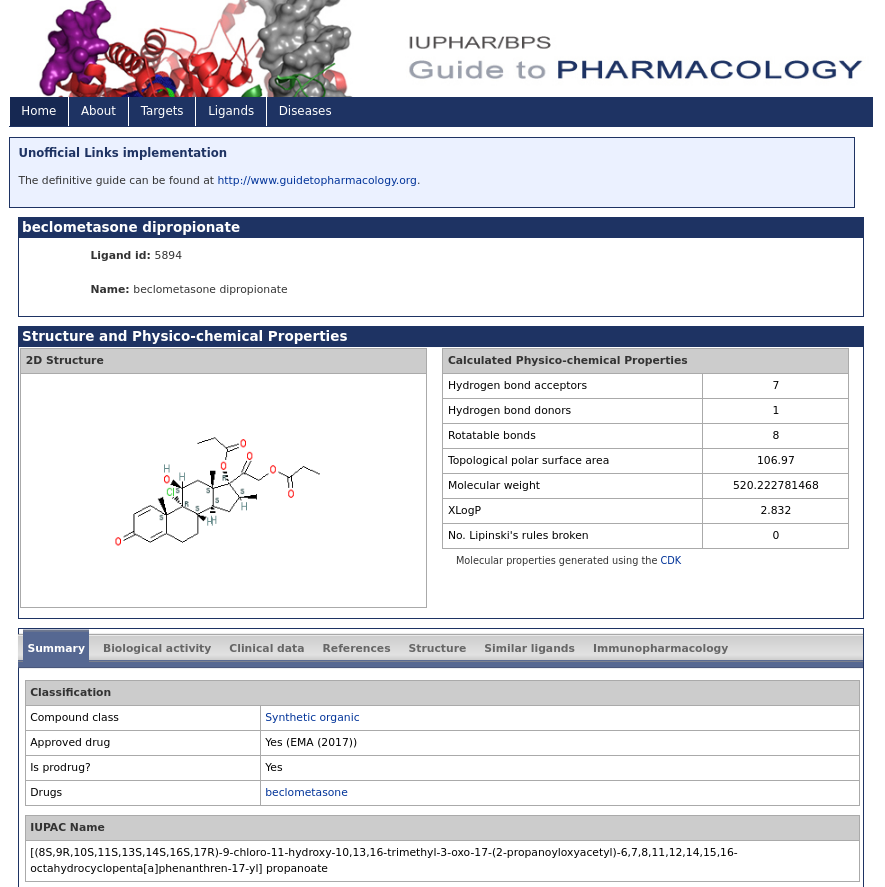}
    \caption{Links reimplementation}
    \label{fig:screenshot:links}
  \end{subfigure}
  \end{center}
  \caption{Screenshots of official GtoPdb application and Links reimplementation}
  \label{fig:reimplementation-screenshots}
\end{figure}

We now turn our attention to our reimplementation of the GtoPdb frontend in
Links. Our reimplementation uses an unaltered copy of the PostgreSQL database
release.

Figure~\ref{fig:reimplementation-screenshots} shows an example page, displaying
a view of the ligand information for \emph{beclometasone dipropionate} extracted from
the database.  Figure~\ref{fig:screenshot:official} shows the official version
of the page, and Figure~\ref{fig:screenshot:links} shows our reimplementation in
Links.  The same underlying information is displayed on each page. The Links
version has some minor differences such as different rounding used for
floating-point numbers, as well as a banner to differentiate it from the
official version.

\subsection{GtoPdb Structure}

The GtoPdb interface consists of nine main data pages:

\begin{description}
  \item[Target List]
    GtoPdb groups pharmacological targets into different categories including
    G protein-coupled receptors, ion channels, nuclear hormone receptors,
    kinases, catalytic receptors, transporters, enzymes, and other protein
    targets. This page links to the family list for each type of target.
  \item[Family List]
    A \emph{family} is a group of related pharmacological targets. The family
    list page displays a hierarchically-ordered tree of families.
  \item[Family Data]
    The family data page provides summaries of each target in the family, and
    links to the more in-depth object data pages.
  \item[Object Data]
    In GtoPdb, an \emph{object} is a pharmacological target such as a receptor.
    The object data page displays all information associated with a target,
    and is the most complex page. The page can render 52 individual properties
    about each object (for example, associated interactions and 3D structures).
  \item[Disease List]
    A list of all diseases in the system.
  \item[Ligand Families]
    A list of ligands, classified into families.
  \item[Ligand List]
    A list of all ligands in the system, with the ability to filter by category.
  \item[Ligand Data]
    Displays data associated with ligands, such as relevant interactions,
    structural information, and a summary of clinical use.
  \item[Disease Data]
    Displays information about a disease, including references to external
    databases, related pharmacological targets, and ligands known to affect the
    disease.
\end{description}

The official implementation contains additional smaller auxiliary data pages and
searching functionality, but the nine pages above are the most prominent and
involved, so we concentrate on these pages for our Links reimplementation.

\subsection{Language-Integrated Query}
In the Java implementation of GtoPdb, all database queries are carried
out using SQL prepared statements. Our first major departure from the
previous implementation is the use of Links's support for
\emph{language-integrated query}.  While language-integrated query is
best known in Microsoft .NET languages such as C\# and
F\#~\citep{meijer:sigmod,Syme06}, it is also available as part of
Links~\citep{Cooper09:linq,LindleyC12:linq}, based on techniques
developed originally in the Kleisli system~\citep{wong2000kleisli}.

Instead of constructing SQL statements directly, language-integrated query
allows database queries to be written as a standard expression in a programming
language. In particular, we use a flavour of language-integrated query pioneered
by~\citet{Trinder91:comprehensions}, who adapts the notion of a list
comprehension (similar to a mathematical set comprehension) to the setting of
relational queries. As an example, consider an SQL expression which retrieves
the names of all ligands which have been approved for use in humans. In SQL, we
might write:

\begin{lstlisting}[language=SQL]
SELECT name FROM ligand WHERE approved
\end{lstlisting}

Given an appropriate Links declaration of the \lstinline+ligand+ table, we can
write the above query as:

\begin{lstlisting}
query { for (l <-- ligand) where (l.approved) [l.name] }
\end{lstlisting}

The query is written as a list comprehension, with elements generated by the
\lstinline+ligand+ table, where the \lstinline+where+ clause filters each
element to only consider those which are approved. Unlike embedded SQL, the
query is also typechecked to ensure the table field names and query results are
used consistently in the program.

Differently from most implementations of language-integrated query,
Links supports efficient \emph{nested queries}.  A nested query is a
query whose result type contains collections nested inside other
collections.  (In contrast, an SQL query always returns a \emph{flat}
table: a collection of records of values of primitive types
such as integers and strings.)  To illustrate, the following nested
query returns records including the ligand name and its set of
synonyms:
\begin{lstlisting}
query { for (l <-- ligand)
          [(name = l.name,
            synonyms = for (l2s <-- ligand2synonym)
                       where (l2s.ligand_id == l.ligand_id)
                       [l2s.synonym])] }
\end{lstlisting}
This query will return a list of records \lstinline{(name,synonyms)}, in which \lstinline{name} is
a string and \lstinline{synonyms} is a list of strings.  A natural,
but inefficient, way to execute such a query is to first retrieve the
set of all ligands (with their names and IDs), and then run one query
per ligand to find its synonyms.  In Links, the above nested query is
instead transformed to two equivalent SQL queries.
It is important to note that the \emph{shredding} technique to implement nested
queries proposed by~\citet{CheneyLW14:shredding} gives a guaranteed upper bound
on the number of SQL queries needed to run a nested query: the upper bound is
the number of occurrences of collections in the query result type, and this is
independent of the number of records returned by a query.

A second useful feature of language-integrated query in Links is that
certain user-defined functions can be used within queries for
convenience, and such functions will be \emph{inlined} to simplify the
query expression to a form that can be translated to SQL.  This relies on query
normalisation~\citep{Cooper09:linq} to generate efficient SQL code directly from
query expressions.  As a simple
example demonstrating both nested queries and user-defined functions
at once, the nested query above can also be written as:
\begin{lstlisting}
fun getSynonyms(id) {
  for (l2s <-- ligand2synonym)
  where (l2s.ligand_id == id)
  [l2s.synonym]
}
query { for (l <-- ligand)
          [(name = l.name,
            synonyms = getSynonyms(l.name))] }
\end{lstlisting}
In this example, this capability makes the query expression longer
(but arguably a bit more readable due to the extra documentation
provided by the function name); however, such functions can also be
\emph{reused} across many query expressions, potentially saving a
great deal of code repetition, and aiding maintenance.

\subsection{Example: Listing Ligands}
GtoPdb provides functionality for listing all ligands in the database, filtered
by category: example categories include approved drugs, or ligands relevant to
either immunopharmacology or malaria pharmacology. Let us consider this page as
an extended example.

Each row in the displayed table shows the ligand's name, unique GtoPdb ID,
synonyms or trademark names, and icons displaying whether the ligand is an
approved drug; contains a radioactive or chemical (e.g. fluorescent) label; is
relevant to immunopharmacology; is relevant to malaria pharmacology; or has an
entry in the protein 3D structure database (PDB).
We can gather this information through the use of a single query expression:

\begin{lstlisting}
query {
  for (l <-- ligand)
    where (ligandFilter(l, filterType))
      [ (id = l.ligand_id, name = l.name, approved = l.approved,
         radioactive = l.radioactive, labelled = l.labelled,
         immuno = l.in_gtip, malaria = l.in_gtmp,
         synonyms =
          for (l2s <-- ligand2synonym)
            where (l2s.ligand_id == l.ligand_id && l2s.display)
              [l2s.synonym],
         hasPDB =
           not(empty(
             for (p <-- pdb_structure)
              where (p.ligand_id == l.ligand_id)
              [ p ])))
      ]
};
\end{lstlisting}

We begin by querying the ligand table. If the ligand matches a given predicate
based on the filter type, then the query produces a record with the required
information. Of particular interest are the \lstinline+synonyms+ and
\lstinline+hasPDB+ fields of the output record, which are not fields in the
ligand table but instead refer to other tables.

The \lstinline+synonyms+ field is a one-to-many relation from ligands to synonyms. As an
example, the common painkiller paracetamol is also known by the trade names
Panadol and Tylenol. The Java implementation gathers the relevant synonyms using
a PostgreSQL view. To express this nested relation in Links, we use its support
for nested queries~\citep{CheneyLW14:shredding}, as explained above, to populate the synonyms
field with a collection of all relevant synonyms.

The \lstinline+hasPDB+ field should be
true if the \lstinline+pdb_structure+ table in the database contains an entry with the same
ligand ID as the current ligand. Note that we can use the native Links functions
\lstinline+not+ and \lstinline+empty+ in Links query code; these are translated to SQL
\lstinline+EXISTS+ constraints.

\subsubsection{Functional Predicates}
Let us revisit how we filter the ligands to display. Some filters are based on
boolean flags in the \lstinline+ligand+ table (for example,
\lstinline+approved+), or on the \lstinline+type+ field,
and others perform more complicated tests. In the Java implementation, such
filtering is implemented by building a query using string concatenation and Java
conditional expressions. For example, to filter all approved drugs, the
implementation uses code like the following:

\begin{lstlisting}
if(type.equalsIgnoreCase("Approved")) {
  query += " WHERE approved IS TRUE ";
} else if (type.equalsIgnoreCase("Synthetic organic")) {
  query += " WHERE type = 'Synthetic organic' "
} ...
\end{lstlisting}

Each ligand type has a case which adds the correct type into the query, chained
as \lstinline+else if+ clauses. Query strings generated in this way could be
ill-formed, leading to failure at runtime (for example, if spaces between
concatenated strings are omitted). Instead, we can take advantage of the fact
that Links is a functional programming language, and define a function that
tests whether a ligand matches a filter. We begin by defining a variant type
(similar to an \lstinline+enum+ in Java) describing each filter:

\begin{lstlisting}
typename Filter = [|
    Approved | SyntheticOrganic | EndogenousPeptide | Immuno | ... |];
\end{lstlisting}

We then define a function called \lstinline+ligandFilter+ that given an entry in
the \lstinline+ligand+ table, and a filter type, returns whether the ligand
matches the filter:

\begin{lstlisting}
fun ligandFilter(ligand, filterType) {
  switch (filterType) {
    case Approved          -> ligand.approved
    case SyntheticOrganic  -> ligand.type == "Synthetic organic"
    case EndogenousPeptide ->
      ligand.type == "Peptide" && isEndogenous(ligand)
    case Immuno            -> ligand.in_gtip
    ...
  }
}
  \end{lstlisting}

Note that the \lstinline+EndogenousPeptide+ case calls another function
\lstinline+isEndogenous+, illustrating that we can use functions to break the
query logic down into smaller parts.  The \lstinline+ligandFilter+ function can
be used directly in the query, and Links correctly inlines it (and
\lstinline+isEndogenous+) so that the eventual SQL query is similar to the one
generated by the Java code. Using the number of lines of code as a rough measure of
complexity, the Java version needs 145 lines of code to filter the list of
ligands, whereas the Links version requires only 54, with the additional
advantage that Links will always generate type-correct queries.

\subsection{Displaying a Data Page}
We have now seen an example of how Links can be used to implement a GtoPdb data
page.  More generally, in both the Java and Links implementations, the process for displaying
each data page is as follows:

\begin{enumerate}
  \item Parse any input arguments to the page request (for example, ligand or
    object ID)
  \item Perform database queries to populate a data model
  \item Parse all text fields in order to obtain a list of any referenced
    scientific literature and relevant ligands
  \item Render the web page content and deliver the response
\end{enumerate}

In the Java implementation, there is a single Java data model used for both the
web interface and the curation tool, and sometimes this means that additional
information is retrieved but not displayed. In the Links implementation,
each page has its own data model based on what is to be displayed to the user, but
the queries and processing code can be reused over different files. The Java
implementation makes use of a data access layer which contains many methods to
populate the model, whereas each Links page begins with a large nested query
followed by a processing phase.

Text fields in GtoPdb may contain references to supporting scientific papers and
crossreferences to ligands. As an example, consider the following excerpt,
detailing comments about the agonist interactions targeting the D$_1$ dopamine
receptor:

\begin{minipage}{\textwidth}
  \vspace{1em}
  \small
\begin{spverbatim}
 Some substituted benzazepines such as SKF-83959 are G-protein biased agonists of the dopamine D<sub>1</sub> receptor and fail to activate &beta;-arrestin recruitment <Reference id=28036/>; their ability to signal through G<sub>q</sub>-mediated pathways has been controversial <Reference id=33435/>.<br><br><Ligand id=6077/>, <Ligand id=9637/> and related compounds exhibit slow dissociation rates from the D<sub>1</sub> receptor.
\end{spverbatim}
  \vspace{1em}
\end{minipage}

References to scientific papers are introduced using \lstinline+Reference+
tags, and references to ligands are introduced using the \lstinline+Ligand+
tag. The text also includes standard HTML tags. The above text would be rendered
as shown below:

\begin{minipage}{\textwidth}
  \vspace{1em}
  \includegraphics[width=\textwidth]{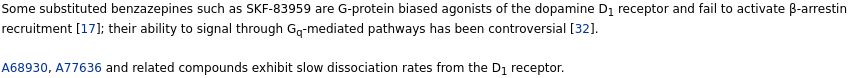}
  \vspace{0.01em}

\end{minipage}

In order to display the text, we need to parse the \lstinline+Reference+ and
\lstinline+Ligand+ tags, so that the required data about the corresponding
references and ligands can be fetched in a subsequent query. This separate pass
allows us to build a sorted, numbered, reference list, rather than storing
references per data page in the database in a more fragile way.

\section{Evaluation}
In this section, we evaluate our Links reimplementation of GtoPdb against the
original.
Firstly, we show a functional correctness analysis, which gives a good degree of
confidence that the Links reimplementation provides an equivalent level of
functionality to the original version.
Secondly, we show a comparative performance analysis, which in particular shows
that Links generates fewer queries, but shows that the relative immaturity of
Links compared to the Java compiler results in longer page build times.

The data generated by our experiments, as well as the code used to generate the
charts, is publicly available on Figshare~\citep{data}.

\subsection{Functional correctness}

The first evaluation criterion is functional correctness: does the Links
reimplementation provide broadly the same functionality as the original?

To this end, we have undertaken an automated functional correctness evaluation,
using a 1300-line Python script. Our approach is to use
Selenium~\citep{HolmesK06:selenium} to load each page and interact with the page
to ensure all details are shown, and then use BeautifulSoup~\citep{beautifulsoup}
to parse the results into a JSON object. The JSON objects for the Links and Java
pages can then be compared directly.

For some less involved pages, such as the family and disease lists, the results
can be compared directly. For larger pages, such as the object and ligand
display pages, parsing the data directly is intractable due to the large number
of different display formats, and subtle differences in rendering across the two
implementations.

However, since GtoPdb only displays a data box if relevant data is present for a
particular entity, and references are contained within the text of each piece of
displayed data, a good approximation is to compare the data headers and
reference list of each page. Although an approximation, this approach was
sufficient for us to identify and fix many omissions and errors, and even helped
identify a minor bug in the official implementation.

As an example, consider the JSON object for \emph{Guanylyl cyclase, 21}:

{\small
\begin{verbatim}
{'basic_info': {'family': 'Nitric oxide (NO)-sensitive '
                          '(soluble) guanylyl cyclase',
                'id': 2897,
                'nomenclature': 'Guanylyl cyclase, 21'},
 'data_headers': ['Activators', 'Database Links',
                  'Endogenous ligand(s)', 'Enzyme Reaction',
                  'Inhibitors', 'Previous and Unofficial Names',
                  'Quaternary Structure: Subunits'],
 'references': [7527671, 9742221, 11242081, 12086987,
                19089334, 28557445, 29859918]}
\end{verbatim}
}

The JSON object contains the basic information for the target, the data headers,
and the Pubmed IDs for each reference in the references list.

We exhaustively verified all main summary pages, namely the family list, disease
list, ligand list, and ligand families pages. (The target list page is
static, so is not of interest for the evaluation). For each data page, we
performed the check on 150 randomly-chosen entity identifiers.

Overall, we managed to validate the vast majority of pages. The one shortcoming
of the Links implementation with respect to functional correctness is that its
parser cannot handle malformed data: in particular, malformed HTML (such as text
including $<$ or $>$ literals instead of the escaped HTML \verb+&lt;+ and
\verb+&gt;+) is unsupported. This leads to some pages failing the references
check, since parsing for some fields fails. However, the number of affected
pages is small: the validation showed errors on two disease data pages, five
family data pages, two object data pages, and no ligand pages. We
manually checked each error to ensure that it was due to malformed HTML.

\subsection{Performance}

The second evaluation criterion is performance. Although we have shown that
Links is capable of reimplementing the main functionality of GtoPdb, and that
our reimplementation displays the same data, it is important to ensure that our
cross-tier methodology is not unacceptably detrimental to performance. We
therefore evaluate our approach on four of the main data pages: the object data
page; the disease data page; and the lists of all diseases and ligands.

We evaluate each page with respect to three dimensions:

\begin{description}
  \item[Query Count]
    The number of queries executed when generating a given page. As the nested
    language-integrated query approach used by Links ensures that query count is
    bounded by the number of collection types in the
    result~\citep{CheneyLW14:shredding}, we would hope that the Links
    implementation would generate substantially fewer queries than
    constructing queries by hand.
  \item[Query Handling Time]
    The amount of time spent processing database queries. In the Links
    implementation, this also includes the time spent normalising the query into
    SQL and parsing the results into Links data structures, in addition to query
    execution itself.
     \item[Page Build Time]
    The amount of time spent building a page on the server, measured from the
    point at which the request is received until the point before the response
    is sent to the client. As Links is an interpreted language (as opposed to
    Java, which runs on a virtual machine incorporating a Just-In-Time (JIT)
    compiler), we would expect page build time to be slower on Links. We show
    results both including and excluding query handling time.
\end{description}

We collect the metrics by adding instrumentation code to the Links interpreter
itself, and by implementing an instrumented version of the
\lstinline+PreparedStatement+ class in the Java code. Measurements were performed
on both the Links and Java versions of the code running locally on a laptop with
an Intel Xeon E-2176M 2.7GHz CPU, and 8GB of RAM. We used the
Python \lstinline+pandas+ library~\citep{pandas} for data processing, and
\lstinline+matplotlib+~\citep{matplotlib} to generate charts.

\subsubsection{Object Data}

\begin{figure}[t]
  \begin{subfigure}[t]{0.45\textwidth}
    \includegraphics[width=0.9\textwidth]{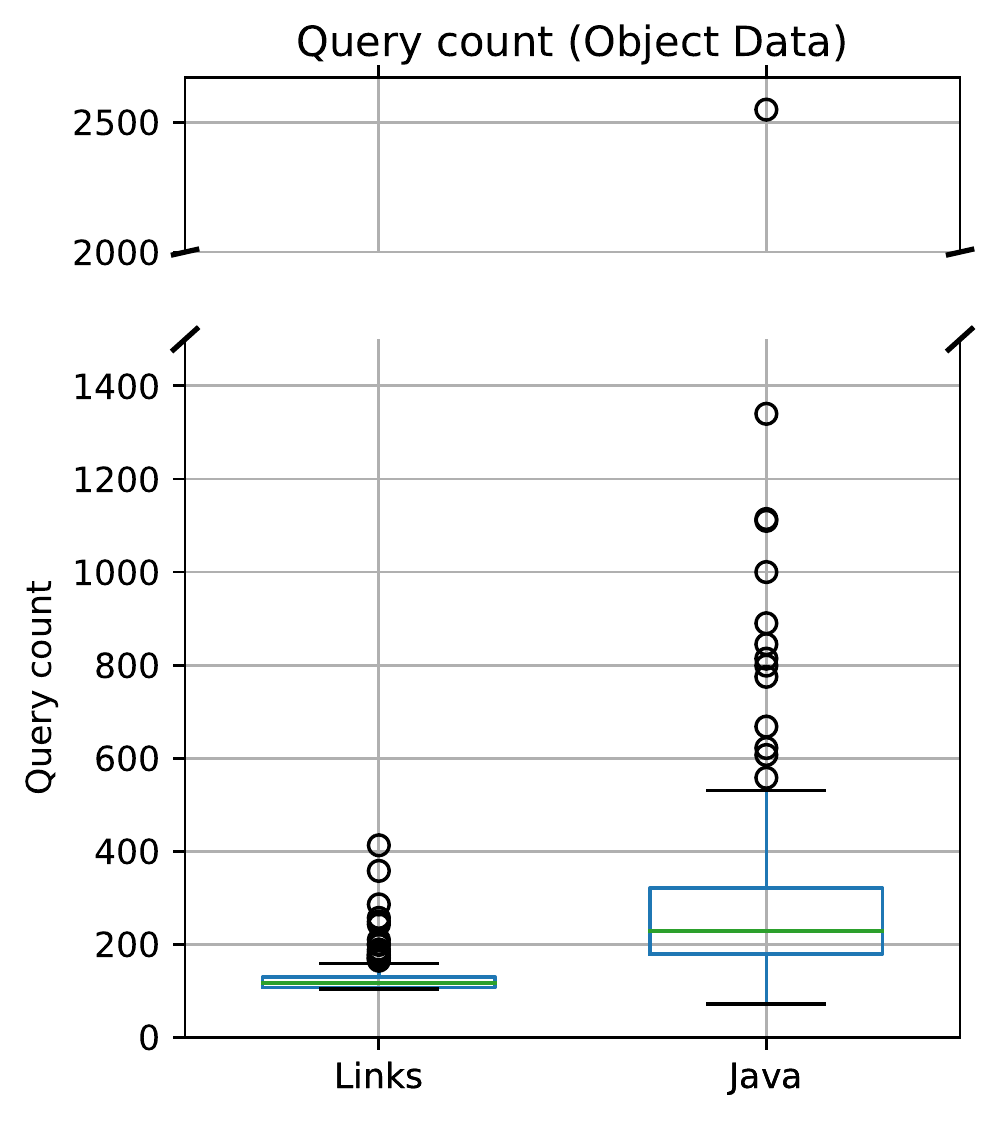}
    \caption{Query count}
    \label{fig:eval:objectdisplay:query-count}
  \end{subfigure}
  \hfill
  \begin{subfigure}[t]{0.45\textwidth}
    \includegraphics[width=0.9\textwidth]{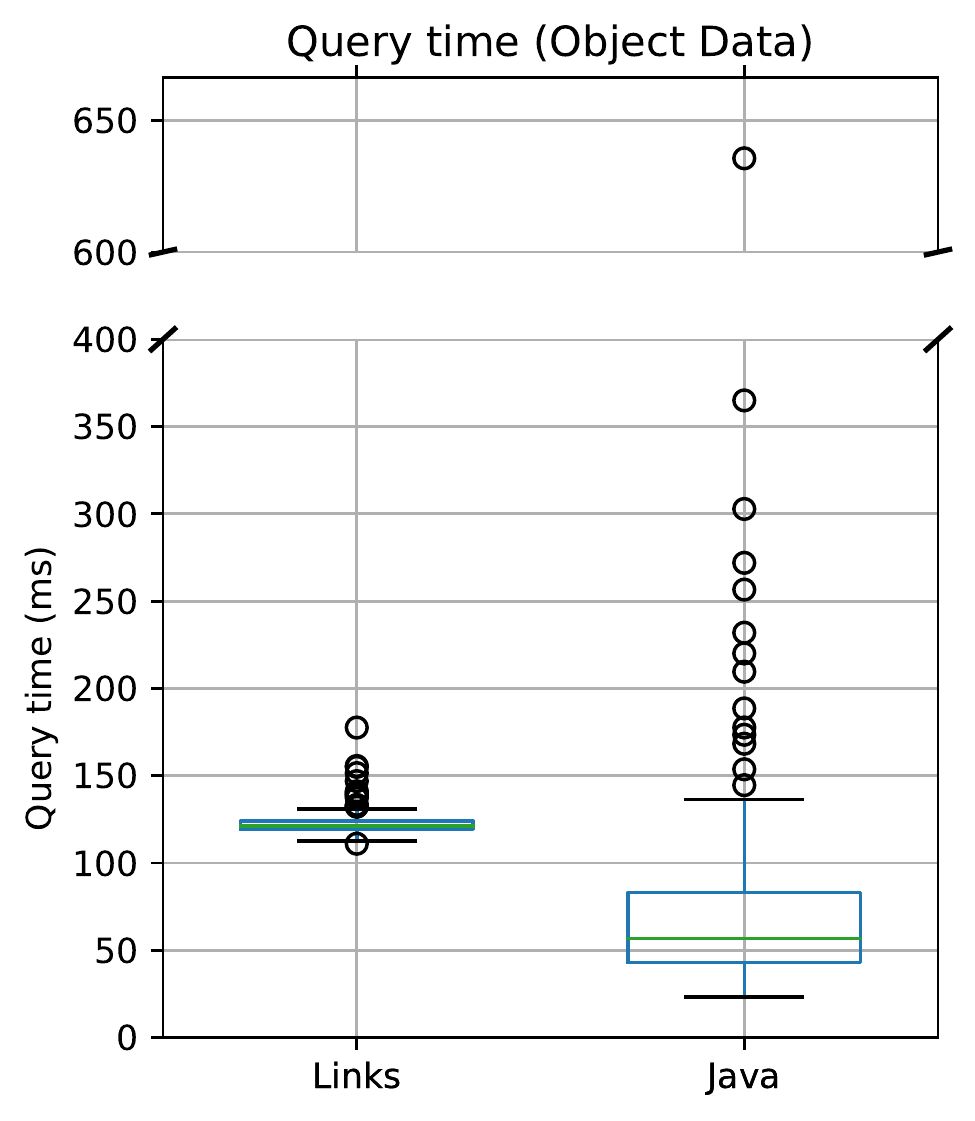}
    \caption{Query handling time}
    \label{fig:eval:objectdisplay:query-time}
  \end{subfigure}

  \medskip

  \begin{subfigure}[t]{0.45\textwidth}
    \includegraphics[width=0.9\textwidth]{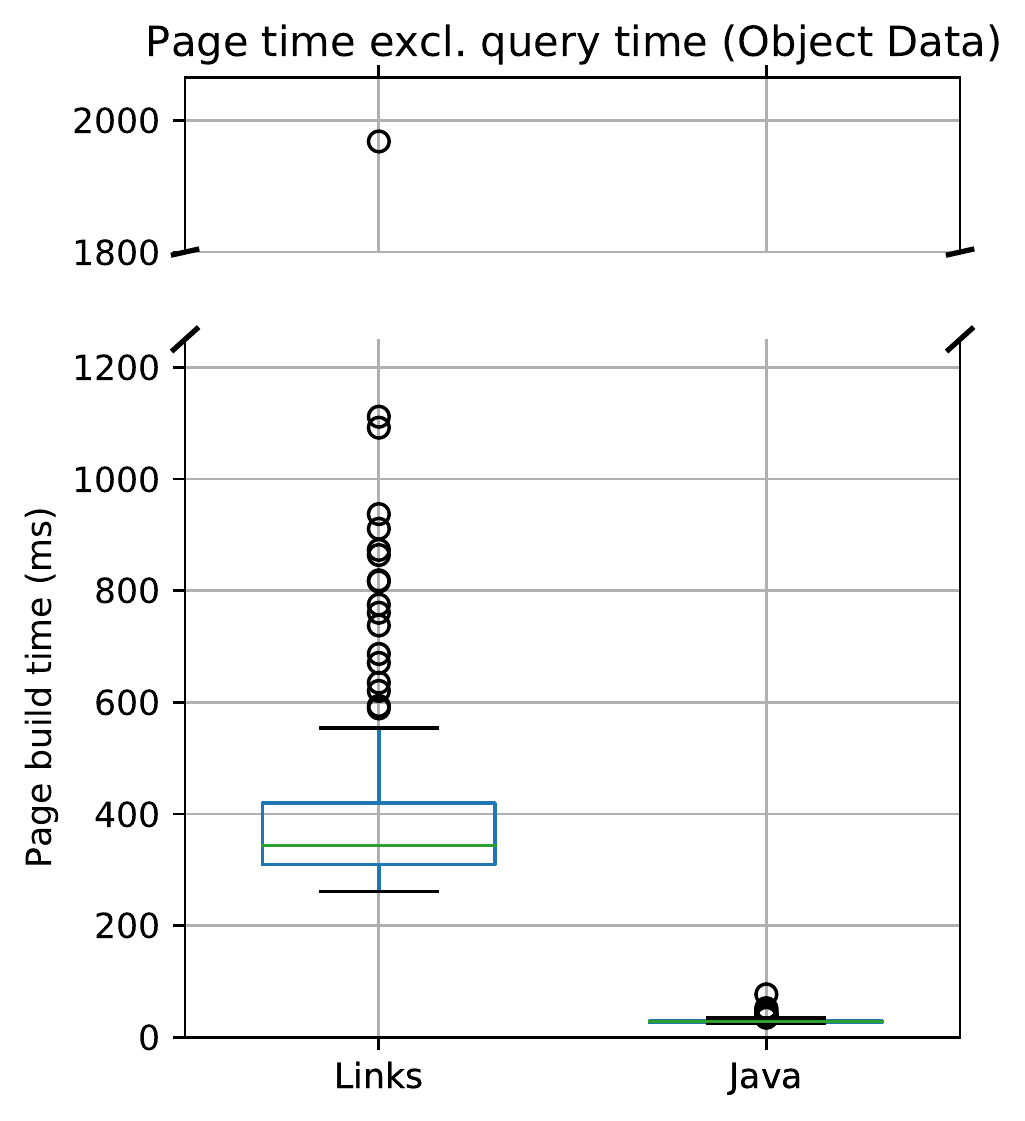}
    \caption{Page build time (excluding queries)}
    \label{fig:eval:objectdisplay:pagebuild-excl-time}
  \end{subfigure}
  \hfill
  \begin{subfigure}[t]{0.45\textwidth}
    \includegraphics[width=0.9\textwidth]{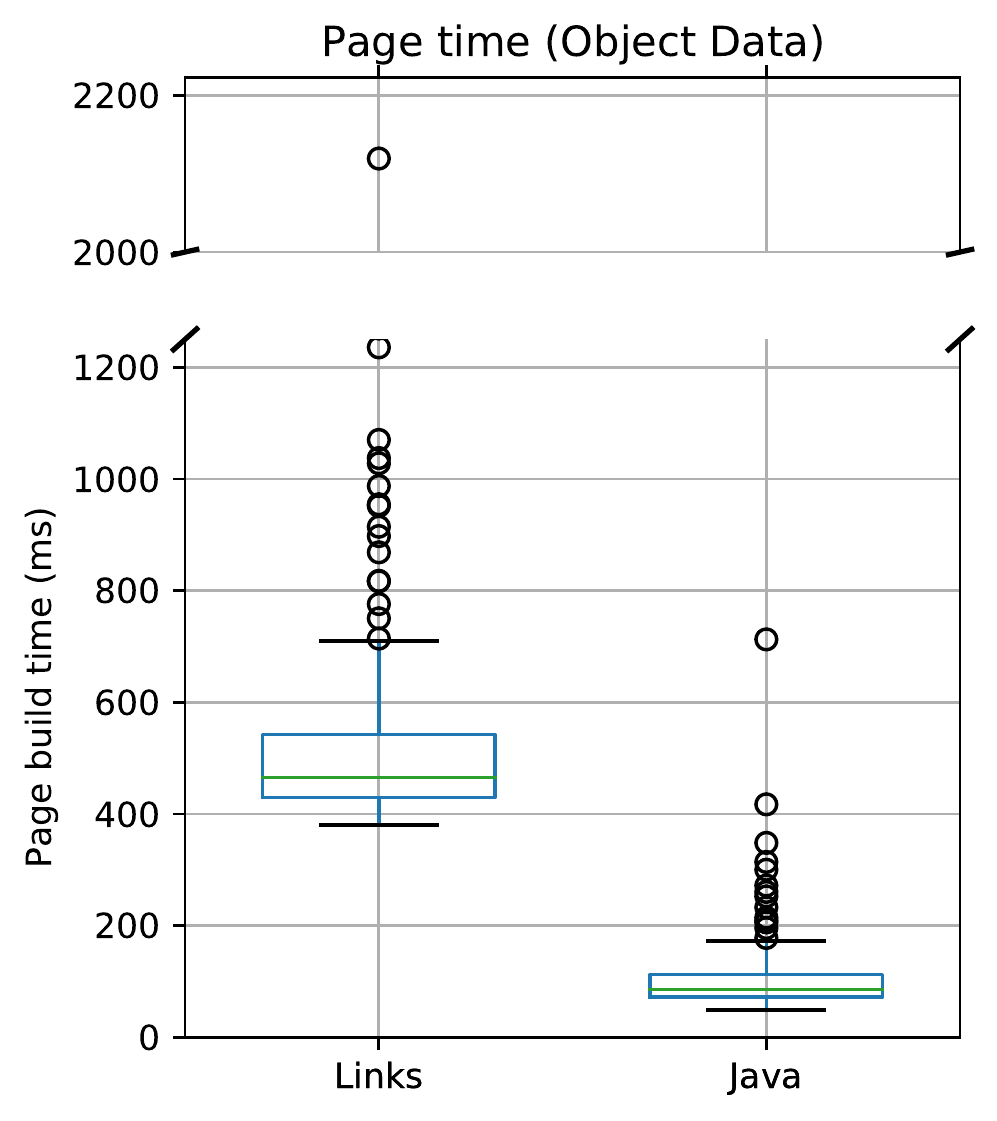}
    \caption{Page build time (including queries)}
    \label{fig:eval:objectdisplay:pagebuild-incl-time}
  \end{subfigure}
  \caption{Experimental evaluation of implementations (Object Data Page)}
  \label{fig:eval:objectdisplay}
\end{figure}

Figure~\ref{fig:eval:objectdisplay} shows box plots detailing the number
of queries, total query handling time, and page build time when displaying the
object data page for 150 randomly-selected object IDs. The data represents
the arithmetic mean of 20 requests for each page.

\paragraph{Query count.}
Figure~\ref{fig:eval:objectdisplay:query-count} shows the results for the number
of queries generated to display the page. As expected due to the use of nested
queries, the Links implementation generates a lower number of queries (median:
117) compared to the Java version (median: 229). Notably, the query count of the
Links implementation is much more predictable, with a standard deviation of
43.38 in the Links implementation, in contrast to 275.15 in the Java
implementation. A smaller standard deviation, and consequently greater
predictability of the query count, is a good indicator that the number of
queries is less dependent on the contents of the data and more uniform across
different data pages.

The maximum query count in the Links implementation is 413, and
the maximum query count in the Java implementation is 2549. The outlier in the
Java implementation is due to object 262 (the Histamine H$_1$ receptor) being
associated with an unusually large number of drug interactions: each interaction
requires 9 database queries, along with additional queries to fetch information
about the ligands associated with the interaction.

\paragraph{Query time.}
Figure~\ref{fig:eval:objectdisplay:query-time} shows the query handling time
for both implementations. The original Java implementation has a better median
query time of 56.72ms compared to 121.31ms in the Links implementation: this
disparity is likely due to a combination of query normalisation and marshalling
the returned values into Links data structures. Nevertheless, again, performance
of the Links implementation is more predictable with a standard deviation of
8.43 compared to 70.83 in the Java implementation.

\paragraph{Page build time.}
Figures~\ref{fig:eval:objectdisplay:pagebuild-excl-time} and
\ref{fig:eval:objectdisplay:pagebuild-incl-time}
show the page build time for
both implementations, excluding and including query handling time respectively.

As expected, the Java version performs substantially better than the Links
version due to the maturity of the Java Virtual Machine and associated Java
ecosystem. Concretely, the median page build time for the Links version is
465.24ms and the median page build time for the Java version is 85.65ms.
Additionally, the performance of the Java version is more predictable, with a
standard deviation of 76.39 compared to 223.34 in the Links implementation.

An additional bottleneck in the Links implementation is the implementation of a
parser which is run on each text field in order to extract inline reference and
ligand IDs contained within text fields. It is likely that improvements in this
part of the code could lead to substantial performance improvements.

\subsubsection{Disease and Ligand Lists}

\begin{figure}
  \begin{subfigure}[b]{0.235\textwidth}
    \includegraphics[width=\textwidth]{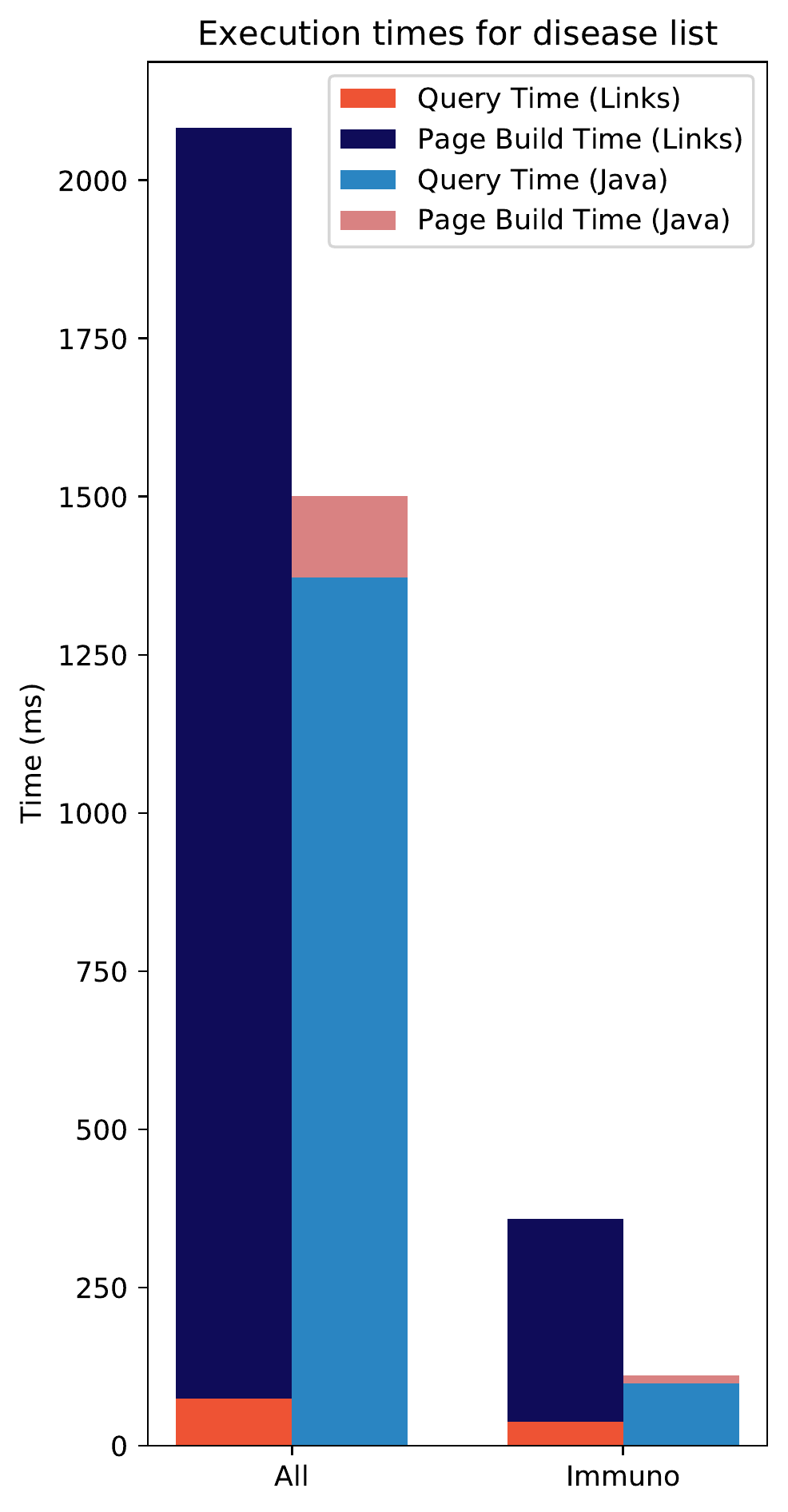}
    \caption{Disease list}
    \label{fig:eval:disease-list}
  \end{subfigure}
    \hfill
  \begin{subfigure}[b]{0.7\textwidth}
    \includegraphics[width=\textwidth]{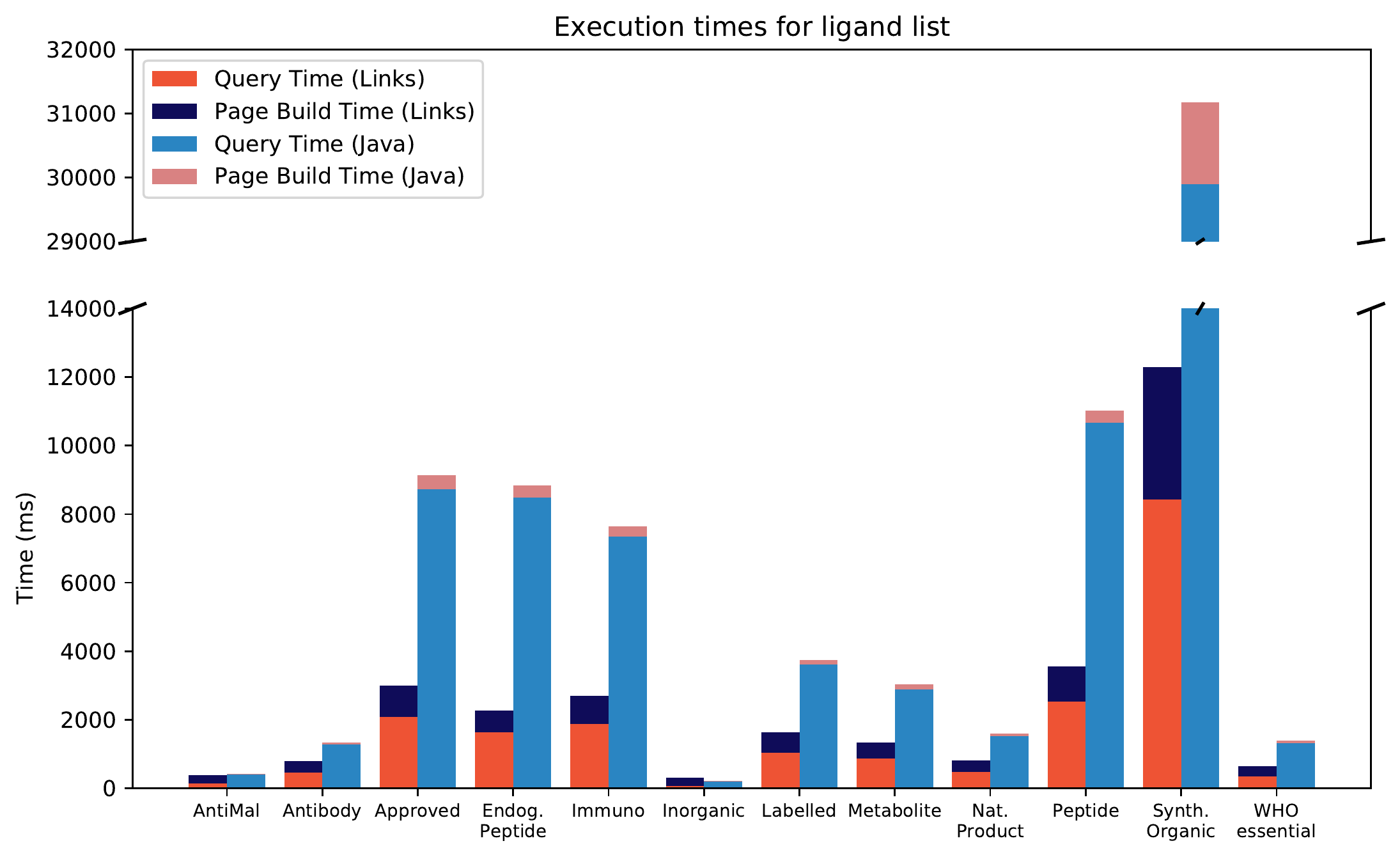}
    \caption{Ligand list}
    \label{fig:eval:ligand-list}
  \end{subfigure}
   \caption{Experimental evaluation of implementations (Disease and Ligand Lists)}
   \label{fig:eval:lists}
 \end{figure}

Figure~\ref{fig:eval:lists} shows the experimental results for the pages listing
all diseases and ligands: the data displayed is again the mean over 20
iterations.
There are two filters for diseases (all diseases, and immuno-relevant diseases),
and twelve filters for ligands, as opposed to thousands of rows of objects and
ligands. Thus, it is possible and instructive to plot the data in a
non-aggregated form.

\paragraph{Query count.}
The Links implementation substantially outperforms that of the Java
implementations: the Links implementations use only two queries to gather the
information required to display the lists, whereas the Java implementation of
the disease list (showing all diseases) requires 8995 queries and the ligand
list (showing approved ligands) requires 30479. The additional number of queries
is because the Java implementation populates a model which contains more
information than is necessary for the page: as an example, the disease list page
generates many queries retrieving links to external databases, but these are
never displayed.

\paragraph{Query time.}
The number of additional queries required in the Java implementations is
reflected in the time spent performing queries. Concretely, the median query time
in the Links implementation was 56.48ms for the disease list and 955.69ms for
the ligand list, compared to 735.97ms for the disease list and 3246.87ms for
the ligand list in the Java implementation.

\paragraph{Page build time.}
As with the object data page, the Links implementation does not perform as
well as the Java implementation when generating the disease list page, due to
the maturity of the underlying technologies. In fact, in spite of the large
number of queries required, the Java implementation outperforms the Links
implementation on the disease list.

However, in spite of the slower page generation time, the much lower query time
in the Links implementation of the ligand list actually allows the Links
implementation to perform better than the Java version on all categories but
one.

\subsubsection{Disease Data}

\begin{figure}[t]
  \begin{subfigure}[t]{0.45\textwidth}
    \includegraphics[width=0.9\textwidth]{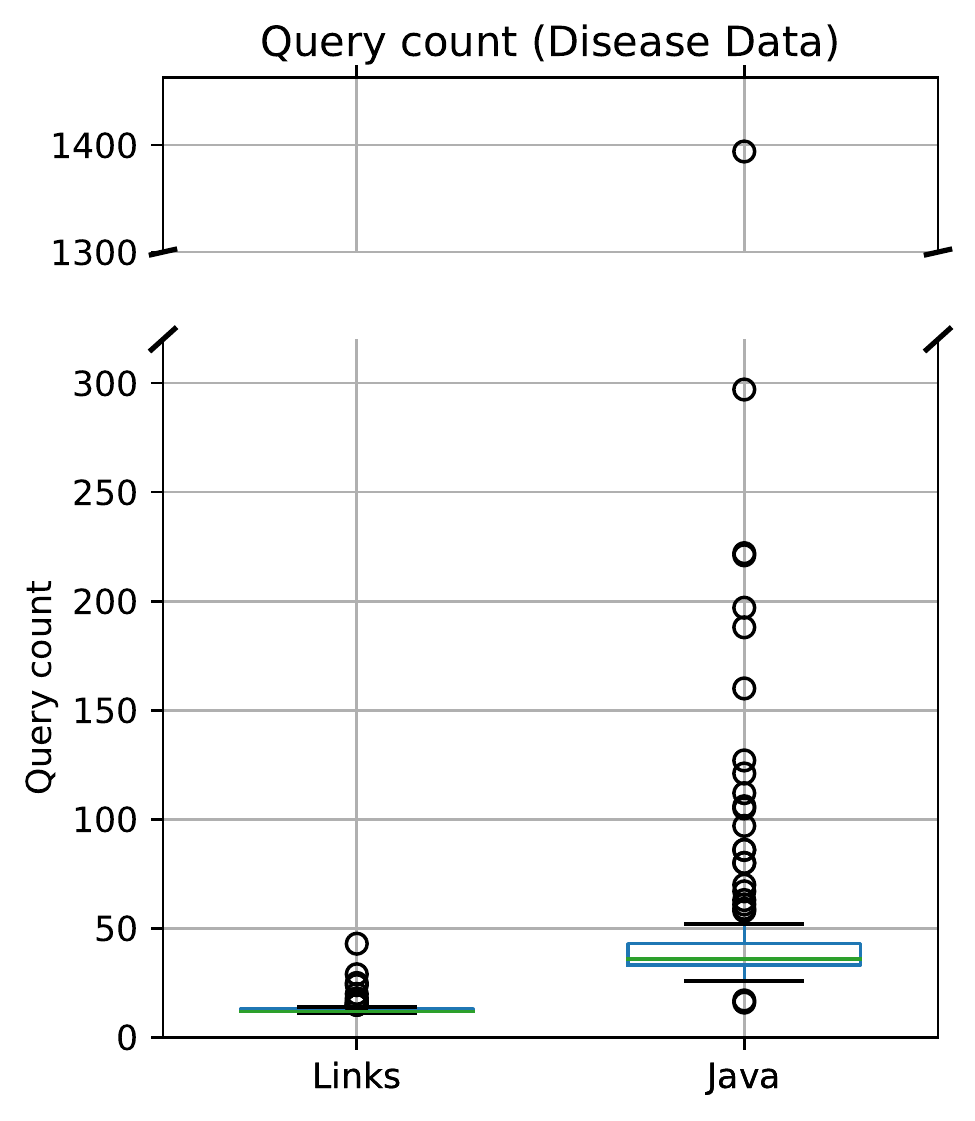}
    \caption{Query count}
    \label{fig:eval:diseasedisplay:query-count}
  \end{subfigure}
  \hfill
  \begin{subfigure}[t]{0.45\textwidth}
    \includegraphics[width=0.9\textwidth]{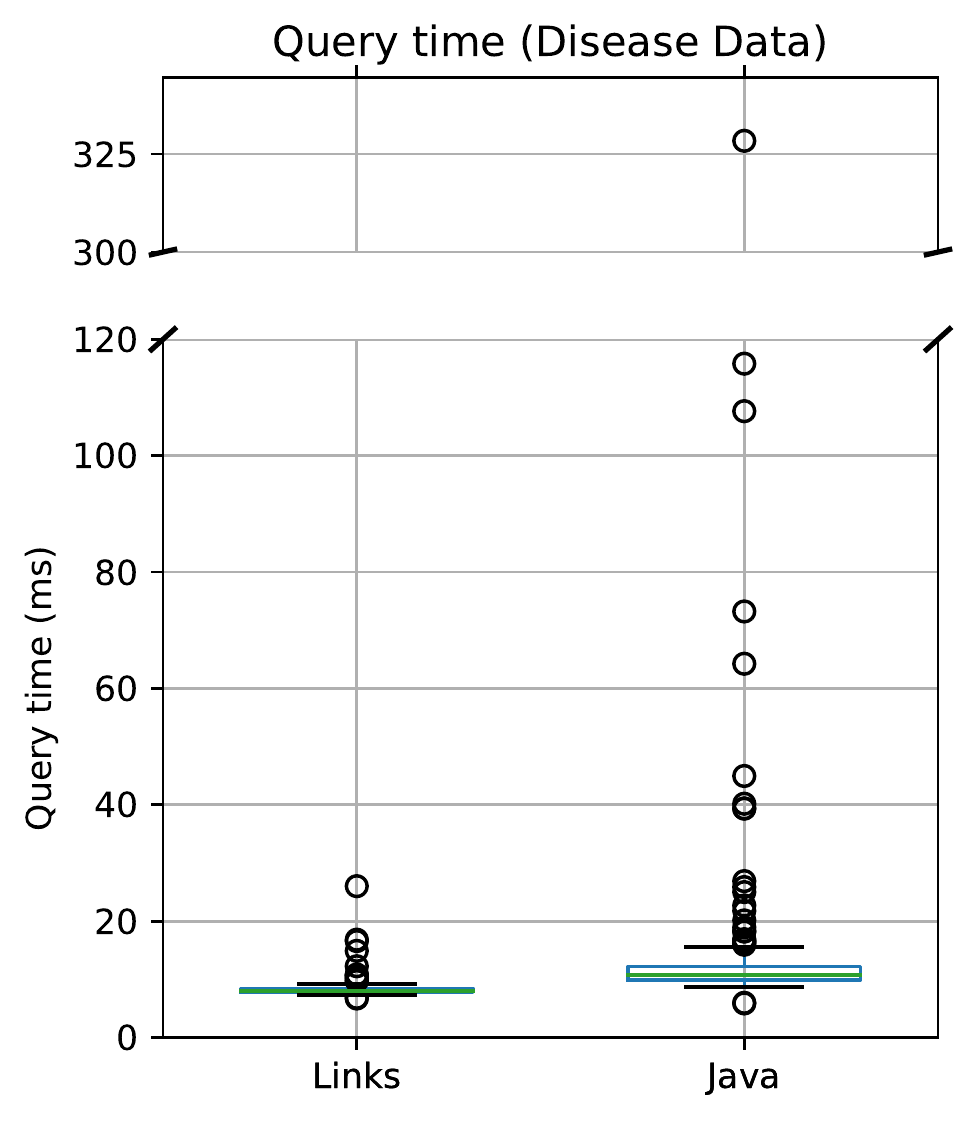}
    \caption{Query handling time}
    \label{fig:eval:diseasedisplay:query-time}
  \end{subfigure}

  \medskip

  \begin{subfigure}[t]{0.45\textwidth}
    \includegraphics[width=0.9\textwidth]{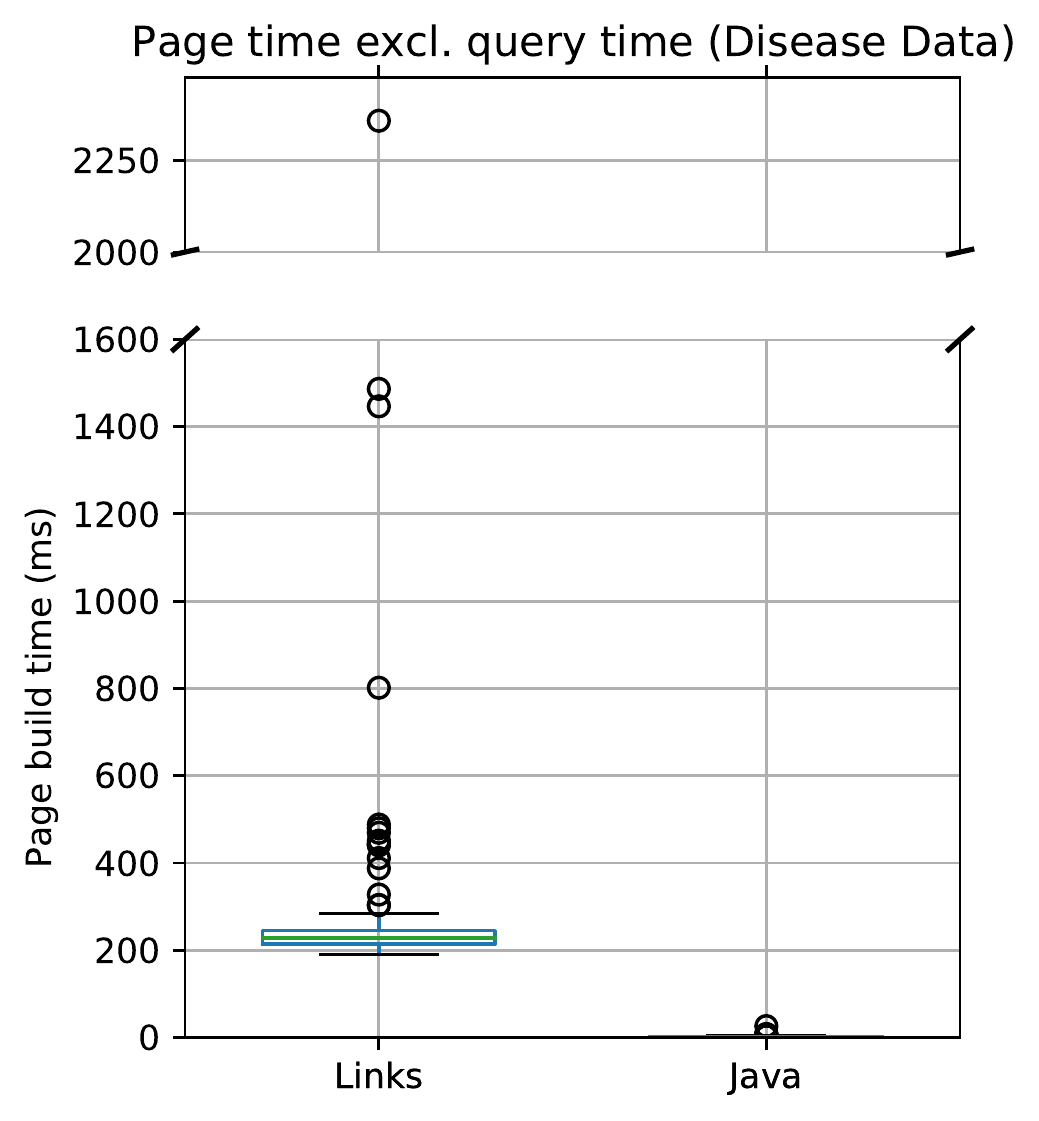}
    \caption{Page build time (Excluding query time)}
    \label{fig:eval:diseasedisplay:pagebuild-excl-time}
  \end{subfigure}
  \hfill
  \begin{subfigure}[t]{0.45\textwidth}
    \includegraphics[width=0.9\textwidth]{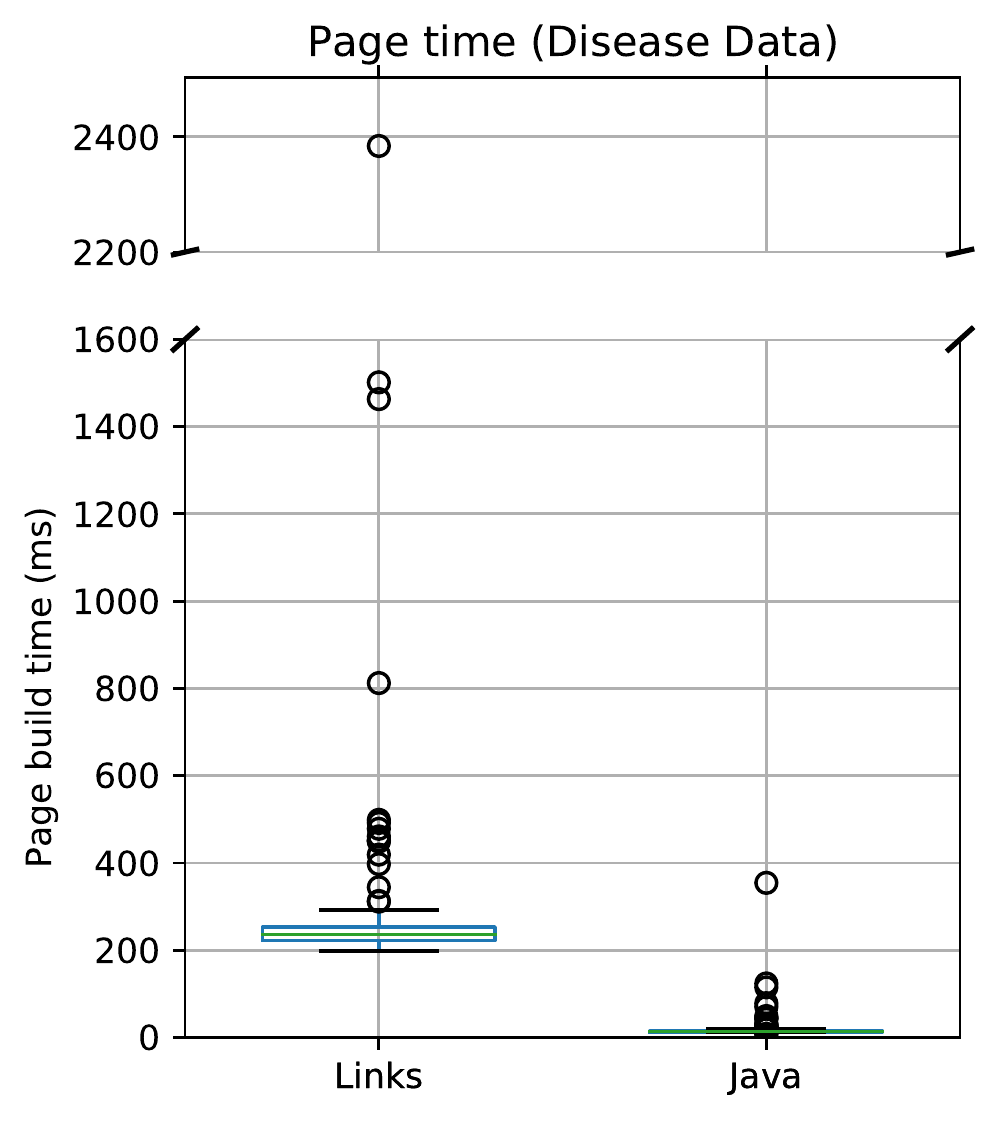}
    \caption{Page build time (Including query time)}
    \label{fig:eval:diseasedisplay:pagebuild-incl-time}
  \end{subfigure}

  \caption{Experimental evaluation of implementations (Disease Data Page)}
  \label{fig:eval:diseasedisplay}
\end{figure}

Figure~\ref{fig:eval:diseasedisplay} shows the results for the disease data
page. Again, the results represent the arithmetic mean over 20
iterations for 150 randomly-selected disease IDs.

The findings are consistent with the previously-reported results, with Links
performing substantially better on query count
(Figure~\ref{fig:eval:diseasedisplay:query-count}) and comparably on query
handling time (Figure~\ref{fig:eval:diseasedisplay:query-time}), but
worse on page build time
(Figure~\ref{fig:eval:diseasedisplay:pagebuild-excl-time}). The outlier for query
count and execution time in the Java implementation is Crohn's disease, which
contains substantially more ligand interactions than the other disease pages
considered. The same disease accounts for the outlier in the Links page build
time, which is due to the necessity of parsing more description fields.

\subsubsection{Summarised Results}

We have looked at two of the list pages and two of the main data display pages.
For completeness, we include summary results of all pages below.

As with the data displayed in previous sections, we repeated all measurements 20
times; the data displayed represents the arithmetic mean of the 20 iterations.
The summary data for data pages represents the median, standard deviation,
minimum, and maximum values for the same 150 randomly-chosen objects as used in
the functional correctness evaluation.

The values for the list data pages represent the same summary statistics but for
the different filters that can be applied to the list (for example, the disease
list allows a user to view all diseases, or just those diseases which are relevant to
immunopharmacology). The number of possible filters is explicitly noted in the
table.

\begin{figure}[t]
{\footnotesize
\begin{tabular}{|c|c|c|c|c|c|c|c|c|}\hline
  & \multicolumn{8}{c}{Query Count} \vline \\ \cline{2-9}
  &
  \multicolumn{4}{c}{Links} \vline &
  \multicolumn{4}{c}{Java} \vline \\ \hline
Page & Median & Std. dev. & Min & Max & Median & Std. dev. & Min & Max \\ \hline
Disease Data & 12 & 3.31 & 11 & 43 & 36 & 116.75 & 16 & 1394 \\ \hline
Disease List $(n = 2)$ & 4 & 0 & 4 & 4 & 4807 & 5922.73 & 619 & 8995 \\ \hline
Family Data & 82 & 126.32 & 49 & 963 & 415 & 1135.82 & 94 & 7634 \\ \hline
Family List $(n = 7)$ & 14 & 24.07 & 3 & 72 & 237 & 492.17 & 69 & 1491 \\ \hline
Ligand Data & 78 & 6.51 & 73 & 115 & 54 & 15.02 & 38 & 126 \\ \hline
Ligand Families $(n = 1)$& 2 & -- & 2 & 2 & 63 & -- & 63 & 63 \\ \hline
Ligand List $(n = 12)$ & 2 & 0 & 2 & 2 & 7757.50 & 23481.39 & 373 & 84712 \\ \hline
Object Data & 117 & 43.38 & 104 & 413 & 229 & 275.15 & 72 & 2549 \\ \hline
\end{tabular}
}
\caption{Summary of query counts}
\label{fig:summary:query-counts}
\end{figure}

\paragraph{Query counts.}
Figure~\ref{fig:summary:query-counts} shows the summary of query counts. As per
our hypothesis, the Links query counts are almost uniformly lower and with
smaller standard deviations.

The one exception is the ligand data page, although this is due to the pages
being architected differently. The ligand data page (see
Figure~\ref{fig:reimplementation-screenshots}) consists of multiple subpages,
accessible via different data tabs. In the Links implementation, data for all
tabs is retrieved on the first page load, and users can navigate the different
tabs without reloading the page.  In the Java implementation, only data for the
relevant tab is loaded, with each subpage requiring a new page load.

\begin{figure}[t]
{\footnotesize
\begin{tabular}{|c|c|c|c|c|c|c|c|c|}\hline
  & \multicolumn{8}{c}{Query Time (ms)} \vline \\ \cline{2-9}
  &
  \multicolumn{4}{c}{Links} \vline &
  \multicolumn{4}{c}{Java} \vline \\ \hline
Page & Median & Std. dev. & Min & Max & Median & Std. dev. & Min & Max \\ \hline
Disease Data & 7.98 & 1.96 & 6.70 & 26.02 & 10.74 & 29.35 & 5.86 & 328.34 \\ \hline
\makecell{Disease List \\$(n = 2)$} & 56.48 & 25.76 & 38.27 & 74.70 & 735.97 & 900.66 & 99.10 & 1372.84 \\ \hline
Family Data & 62.75 & 28.19 & 51.23 & 253.36 & 95.32 & 280.91 & 29.59 & 1883.68 \\ \hline
\makecell{Family List \\$(n = 7)$} & 5.05 & 8.56 & 1.16 & 25.96 & 112.32 & 205.98 & 32.72 & 630.99 \\ \hline
Ligand Data & 121.71 & 335.79 & 108.65 & 2605.65 & 40.87 & 6.23 & 38.30 & 62.60 \\ \hline
\makecell{Ligand Families \\$(n = 1)$} & 0.87 & -- & 0.87 & 0.87 & 33.68 & -- & 33.68 & 33.68 \\ \hline
\makecell{Ligand List \\$(n = 12)$} & 955.69 & 2282.65 & 69.94 & 8434.76 & 3246.87 & 8264.99 & 205.43 & 29892.90 \\ \hline
Object Data & 121.31 & 8.43 & 111.02 & 177.57 & 56.72 & 70.83 & 23.09 & 635.54 \\ \hline
\end{tabular}
}
\caption{Summary of query handling times}
\label{fig:summary:query-times}
\end{figure}

\paragraph{Query times.}

The picture is similarly good with query handling times, shown in
Figure~\ref{fig:summary:query-times}; the median query times are generally
lower in the Links reimplementation (with the exceptions of the ligand data and
object data pages), and the standard deviations are generally far lower,
indicating greater predictability.

\begin{figure}[t]
{\footnotesize
\begin{tabular}{|c|c|c|c|c|c|c|c|c|}\hline
  & \multicolumn{8}{c}{Page Build Time (ms)} \vline \\ \cline{2-9}
  &
  \multicolumn{4}{c}{Links} \vline &
  \multicolumn{4}{c}{Java} \vline \\ \hline
Page & Median & Std. dev. & Min & Max & Median & Std. dev. & Min & Max \\ \hline
Disease Data & 235.70 & 234.82 & 197.76 & 2384.15 & 13.37 & 31.38 & 7.83 & 354.47 \\ \hline
\makecell{Disease List \\$(n = 2)$} & 1221.12 & 1218.97 & 359.18 & 2083.07 & 805.94 & 983.48 & 110.52 & 1501.37 \\ \hline
Family Data & 431.81 & 479.93 & 309.80 & 4070.39 & 104.47 & 299.13 & 33.08 & 2002.07 \\ \hline
\makecell{Family List \\ $(n = 7)$ }& 223.63 & 23.35 & 208.44 & 279.54 & 119.73 & 217.22 & 36.42 & 669.33 \\ \hline
Ligand Data & 439.45 & 421.70 & 346.28 & 3507.80 & 44.05 & 6.37 & 40.91 & 66.74 \\ \hline
\makecell{Ligand Families \\ $(n = 1)$} & 231.35 & -- & 231.35 & 231.35 & 36.89 & -- & 36.89 & 36.89 \\ \hline
\makecell{Ligand List \\ $(n = 12)$} & 1483.64 & 3274.54 & 314.46 & 12292.94 & 3392.51 & 8610.89 & 214.41 & 31170.69 \\ \hline
Object Data & 465.24 & 223.34 & 380.51 & 2119.50 & 85.65 & 76.39 & 49.00 & 712.82 \\ \hline
\end{tabular}
}
\caption{Summary of page build times}
\label{fig:summary:page-times}
\end{figure}

\paragraph{Page build times.}
As discussed previously, the Links page build times
(Figure~\ref{fig:summary:page-times}) are generally worse than the Java page
build times (with the exception of the ligand list page), however the
median page build times are generally still well under a second, so we would
argue that the performance detriment is not prohibitive, especially given the
greater query safety guarantees and performance.

\section{Related work}

Our case study uses Links.  There are other cross-tier languages,
including Hop~\citep{SerranoGL06:hop}, Ur/Web~\citep{Chlipala15:urweb}, and Eliom~\citep{RadannePVB16:eliom}.  To the
best of our knowledge, none of them has been used to implement curated
databases.

Language-integrated query support is now being considered in several
languages, for example including the Quill library for
Scala\footnote{Quill, https://getquill.io/}.
We mentioned Links's support for nested queries as an important
advantage of using Links for implementing GtoPdb.  Similar techniques
offering similar guarantees have been proposed by~\citet{grust10vldb},
with the most recent step in this line of work being a
language-integrated query library for Haskell called DSH~\citep{dsh}.
It might be interesting to conduct a similar case study implementing
GtoPdb in Haskell using DSH, and compare with the Links version;
alternatively it may also be worthwhile to develop a Java or Scala
implementation of language-integrated query that supports nested
queries, that could be used natively with GtoPdb or other Java-based
systems.

\section{Conclusion and Future Work}
In this work, we have produced the first real-world case study of a curated
scientific database, the IUPHAR/BPS Guide to Pharmacology, in a cross-tier
functional programming language. Our approach leverages language-integrated
query, which makes it possible to write type-safe queries instead of manually
constructing SQL.

GtoPdb is a substantial curated database, built over a period of 16 years.
The current Links implementation runs on an unmodified version of the GtoPdb
database release, with the 9 major data pages implemented.  The codebase of our
Links case study currently stands at 13812 lines of code after around 5 months
of effort by the first author, who had previous experience with Links. While it
may be tempting to attempt a direct comparison on lines of code for each page,
such a comparison would be unreliable due to the difference in the structure of
the two applications.

Finally, we have performed an automated functional correctness check, and
conducted a performance evaluation and shown that the use of Links does not
impose unacceptable overheads; indeed, the use of language-integrated query and
nested queries results in lower query counts and comparable time spent handling
queries in general.
The evaluation has shown us that cross-tier programming languages have
\emph{equivalent} functionality to traditional tiered systems, however it is
also worth noting that writing code in a single language and allowing early
static error detection increases the robustness of the technical infrastructure
in the long term.

\paragraph{Future work.}
Our experience shows that Links is up to the task of implementing web
application front-ends for curated databases, which is prerequisite to our goal
of language support for data curation. We have already begun implementing the
GtoPdb curation interface in Links, in particular using \emph{relational
lenses}~\citep{BohannonPV06:rel-lenses, HornPC18:incr-rel-lenses}
to integrate curation activities more tightly with web-based graphical user
interfaces~\citep{HornFC20:links-lens}.

Our next step is to design and implement language features which will aid
curation, such as archiving, inspired by the work
of~\citet{BunemanKTT04:archiving}.

Our work has concentrated on relational databases; we also plan to investigate
language-integrated query for NoSQL databases,
allowing us to implement case studies for a wider range of databases.  This
requires first adapting existing work on language-integrated query from
relational to NoSQL data models and query languages.

\section*{Acknowledgements}
We thank the IDCC and IJDC reviewers for their helpful comments.
This work was supported by ERC Consolidator Grant Skye (grant no. 682315), and
an ISCF Metrology Fellowship grant provided by the UK government's Department
for Business, Energy and Industrial Strategy (BEIS).
Fowler was also supported by EPSRC Grant EP/K034413/1 (ABCD).
The IUPHAR/BPS Guide to
PHARMACOLOGY is supported by the International Union of Basic and Clinical
Pharmacology (IUPHAR), and the British Pharmacological Society (BPS).

\bibliographystyle{apa}
\bibliography{bibliography}

\end{document}

%% file: preamble.tex
\usepackage[sort&compress]{natbib}
\usepackage{subcaption}
\usepackage{listings}
\usepackage{xcolor}
\usepackage{spverbatim}
\usepackage{wrapfig}
\usepackage{beramono}[scale=0.7]
\usepackage{multirow}
\usepackage{makecell}
\usepackage{graphicx}
\usepackage{mathptmx}
\usepackage[T1]{fontenc}

\lstdefinelanguage{Links}{%
  morekeywords={typename, fun, linfun, op, var, if, this, true, false, else, case, switch, handle,
    handler, shallowhandler, open, do, sig, new, send, receive, spawnAt, spawn,
module, request, accept, try, as, otherwise, catch, offer, select, raise,
fork, spawnClient, cancel, catch, page, close, Any, Type, Unl, forall, vdom,
query, where, for},%
  sensitive=t, %
  comment=[l]{\#\ },%
  escapeinside={(*}{*)},%
  morestring=[d]{"},%
  keywordstyle=\color{blue},
  showstringspaces=false
 }

\usepackage{array,ragged2e}
\newcolumntype{P}[1]{>{\RaggedRight\arraybackslash}p{#1}}